\documentclass[submission,copyright,creativecommons]{eptcs}
\providecommand{\event}{ESSS 2015} % Name of the event you are submitting to
\usepackage{breakurl}

\usepackage[utf8]{inputenc}

\usepackage{amsmath}
\usepackage{amsfonts}
\usepackage{amssymb}
\usepackage{amsthm}

\theoremstyle{definition}
\newtheorem{definition}{Definition}

\usepackage{graphicx}
\usepackage{enumerate}

\title{Automatic Generation of Minimal Cut Sets}
\author{Sentot Kromodimoeljo and Peter A. Lindsay
\institute{School of IT\&EE, The University of Queensland, St Lucia Qld 4072, Australia}
}

\def\titlerunning{Automatic Generation of Minimal Cut Sets}
\def\authorrunning{S. Kromodimoeljo \& P.A. Lindsay}

\begin{document}
\maketitle

\begin{abstract}
A cut set is a collection of component failure modes that could lead to a system failure.
Cut Set Analysis (CSA) is applied to critical systems to identify and rank system vulnerabilities at design time.
Model checking tools have been used to automate the generation of minimal cut sets but are generally based on checking reachability of system failure states.
This paper describes a new approach to CSA using a Linear Temporal Logic (LTL)
model checker called BT Analyser that supports the generation of multiple counterexamples. 
The approach enables a broader class of system failures to be analysed, by generalising from failure state formulae to failure behaviours expressed in LTL.
The traditional approach to CSA using model checking
requires the model or system failure to be modified, usually by hand, to eliminate already-discovered cut sets, and the model checker to be rerun, at each step. By contrast, the new approach works incrementally and fully automatically, thereby removing the tedious and error-prone manual process and resulting in significantly reduced computation time. This in turn enables larger models to be checked. 
Two different strategies for using BT Analyser for CSA are presented. 
There is generally no single best strategy for model checking: their relative efficiency depends on the model and property being analysed. Comparative results are given for the A320 hydraulics case study in the Behavior Tree modelling language. 

\end{abstract}

\section{Introduction}

A {\em cut set} is a collection of component failures that could lead to a system failure. 
A cut set is {\em minimal} if none of its proper subsets are themselves cut sets. 
Cut Set Analysis (CSA) is the discovery of a complete set of minimal cut sets (MCSs) for given system failure modes. 
CSA, or an equivalent method such as Fault Tree Analysis (FTA), is typically mandated by standards for critical systems (e.g. \cite{iec61508}) to identify and rank system vulnerabilities at design time.

Traditionally, a system failure mode, or {\it top event} to give it its technical name,
has to be identified before CSA can proceed.
Model checkers are often used to automate CSA when the top event can be characterised by a state
formula. 
CSA proceeds by determining if the top event is reachable for various component failure-mode combinations (the cut sets). Of particular interest are failure combinations that are minimal (the MCSs). 
For some modelling notations, CSA has been fully automated \cite{ake99,bie02,boz07}.

Rae and Lindsay \cite{rae04} generalised the characterisation of system failure in FTA to violation of a temporal property rather than simply a state formula. 
This approach enables a broader class of system failures to be examined, such as state changes under circumstances when states shouldn't change, and action sequences occurring in an undesirable order. 
While such properties can sometimes be captured by adding an observer automaton to the model, it is often more natural to state the property as a temporal property, and desirable not to modify the model \cite{lin12}. 
And in other examples the failures themselves are actually behaviors rather than events or conditions: for example Cerone {\em et al} \cite{cerone08} classify human failures by repeated patterns of behaviour, in a highly interleaved cognitive task where it is impossible to say exactly when the failure occurred. 

Lindsay {\em et al} \cite{lin12b} developed a semi-automated approach to CSA with this richer notion of failures, using Behavior Trees (BT) to model the system with potential component failure-behaviours injected. The SAL model checker \cite{mou04} was used iteratively to identify MCSs, with the temporal property reformulated manually at each step to ignore previously discovered MCSs. The manual steps were tedious and error-prone. Moreover, in order to generate a complete CSA, the model structure needed to be ``flat'', in the sense that if failure event $C1$ is preceded by failure event $C2$ in some path, then there also exists
a path in which $C1$ occurs but $C2$ does not. (This is a stronger assumption than simply that failures are independent.)

This paper describes a completely new approach to automation of CSA, using a new model checker
called BT Analyser, in which search for counterexamples can be directed \cite{kro14}.
As well as avoiding the need for manual steps and the assumption about non-dependence of component failures, the checker is significantly more efficient than the one used in \cite{lin12b}, with capabilities that go well beyond CSA. 
The automation takes advantage of some novel features of BT Analyser including the use of
{\em cycle constraints} for counterexample generation, {\em global constraints}, and
incremental analysis,
resulting in an efficient tool for CSA. This paper also compares alternative
strategies for the automatic generation of minimal cut sets to illustrate the virtues of the novel features of
BT Analyser.
The approach is illustrated on a BT model to enable comparison with previous approaches \cite{lin12b} but it is applicable to any modelling notation that can be translated to a finite state transition system.

\begin{itemize}
\item Section \ref{sec:ta} describes the modelling framework and introduces CSA terminology and concepts.
\item Section \ref{sec:mc} describes the LTL model checker for Behavior Trees.
\item Section \ref{sec:generating} describes the techniques for generating minimal cut sets.
\begin{itemize}
\item Section \ref{sec:extracting} describes how a cut set is extracted from a counterexample path.
\item Section \ref{sec:verifying} describes alternative approaches for verifying the minimality of a cut set.
\item Section \ref{sec:strategies} discusses two strategies for automating the generation of all minimal cut sets.
\end{itemize}
\item Section \ref{sec:experiments} presents results of experiments on applying the techniques to the Airbus A320
hydraulic system case study used in \cite{bie02} and \cite{lin12b}.
\item Section \ref{sec:dis} discusses application of the approach to other modelling notations. 
\item Sections \ref{sec:relwork}, \ref{sec:concl} and \ref{sec:future} discusses related work,
provides a summary and conclusion, and discusses future work, respectively.
\end{itemize}

\section{Terminology and assumptions}
\label{sec:ta}

The core of BT Analyser works for a very general class of modelling languages, including asynchronous finite-state systems with interleaving semantics \cite{kro14}. It does so by using a modelling framework based on typed multi-variable state transition systems in which the effect of each transition is deterministic (but the choice of transition is non-deterministic)
and the truth value of each atomic proposition in a state is determinable. 
Modelling languages that can be translated into this framework include state machines, Behavior Trees and activity diagrams. 

In what follows we assume the component failure modes of interest have already been injected into the model of system behaviours: that is, the analyst has determined all of the component failure modes that are going to be analysed and has included events
(called {\em basic events} in CSA and FTA) corresponding to occurrences of component failures. The effects of
a component failure may need to be reflected in the system behaviour. However, we do not assume that a failed component
remains failed.
As usual for causal analysis techniques such as CSA and FTA, we assume component failures are independent:
common mode failures are analysed after CSA \cite{lev95}.

For the purpose of CSA, we assume that basic events are encoded using special Boolean state variables,
called {\em basic event flags} below.
Each of these state variables represents the occurrence of the corresponding event.
Initially, the state variable must have the value $\mathit{false}$. When the corresponding event occurs,
the value of the state variable must change to $\mathit{true}$ and no behaviour
can change the value back to $\mathit{false}$. If the analyst's modelling notation does not support
this, it can be easily added during translation to BT Analyser's modelling framework.
Note that it is the event that cannot be undone, not the failure of a component.

Given a system safety property $\mathit{Safe}$ expressed in Linear Temporal Logic (LTL) \cite{pnu77},
the model checker investigates whether there is a {\em counterexample} to $\mathit{Safe}$: that is, a (possibly infinite) sequence $\pi$ of transitions through the model which satisfies $\lnot \mathit{Safe}$. (In fact the model checker looks for paths with a finite prefix $p$ and an infinitely repeated subpath $c$, but if there is a counterexample at all then there is always one in the above form.)
Note that {\em safety property} in this paper means the formalisation of a system safety requirement,
rather than the narrow technical sense of $\mathbf{G} \lnot P$ for a state condition $P$.

With the top event replaced by ``a violation of system safety property'', a cut set is
the set of basic events that occur in a behaviour that violates the system safety property
(a counterexample path for $\mathit{Safe}$). The set of basic events that occur in
a counterexample path can be extracted from the cycle part of the counterexample
path: they are exactly the basic events whose corresponding basic event flags have
the value $\mathit{true}$ in the cycle (recall that basic event flags
can only change value from $\mathit{false}$ to $\mathit{true}$, so in a cycle they
must maintain their values).

Two different strategies are given in Section \ref{sec:generating} below for generating counterexamples corresponding to MCSs.

\section{BT Analyser: A Symbolic LTL Model Checker}
\label{sec:mc}

BT Analyser is a symbolic LTL model checker for Behavior Trees. It implements
novel techniques for LTL model checking and counterexample generation described in \cite{kro14}.
Although BT Analyser only accepts the Behavior Tree (BT) notation as the modelling
notation, the core of the model checker is notation-independent.

Novel features of BT Analyser include:
\begin{itemize}
\item Directed counterexample path generation with cycle constraints and global constraints.
\item Computation of fair states with global constraints.
\item On-the-fly symbolic LTL model checking as well as the traditional fixpoint approach to symbolic model checking.
\item Incremental analysis.
\end{itemize}

BT Analyser is a symbolic model checker, operating on sets of states characterised using
propositions rather than individual states. A {\em symbolic state} is a proposition characterising a set of states.

BT Analyser's framework uses {\em elementary blocks} as abstract transitions.
The effect of an elementary block transition is deterministic;
the non-determinism is in choosing an elementary block to transition, when several are enabled.

For each elementary block $b$, let $f_b$ be the function that computes the symbolic image under
the transition (sub-)relation for $b$ and let $r_b$ compute the symbolic preimage.
If $B$ is the set of elementary blocks for the system modelled, then the function $f_B$
for computing the symbolic image under the transition relation for the system is
defined as
\begin{displaymath}
f_B(S) \triangleq \bigvee_{b \in B} f_b(S).
\end{displaymath}
Similarly for the preimage,
\begin{displaymath}
r_B(S) \triangleq \bigvee_{b \in B} r_b(S).
\end{displaymath}
Propositional operations are performed using ordered binary decision diagrams (OBDDs) \cite{bry86}.

Following model checking convention, let the negation of the system safety behaviour being analysed be denoted $\varphi$.
During model checking, a {\em tableau} (decision graph) for $\varphi$ is superimposed on the model producing
an augmented model (see \cite{kro14}). The
image and preimage functions for the augmented model are denoted $f'_b$ and $r'_b$ for
an elementary block $b$ and $f'_B$ and $r'_B$ for the entire system.

\begin{definition}[Symbolic Counterexample Path]
A symbolic counterexample path has the form of a triple:
\begin{displaymath}
  (I_{\pi}, p_{\pi}, s_{\pi})
\end{displaymath}
where $I_{\pi}$ characterises a set of initial states,
prefix $p_{\pi}$ is a possibly empty finite sequence of elementary blocks,
and cycle $s_{\pi}$ is a non-empty finite sequences of elementary blocks that is repeated forever.
A symbolic counterexample path represents a set of paths in the model that satisfy $\varphi$
(the paths share the abstract transitions and all satisfy $\varphi$).
\end{definition}
The intermediate symbolic states in a symbolic path can be computed from $I_{\pi}$ and the image functions for the transitions
(the elementary blocks) in the symbolic path.

Directed counterexample path generation allows BT Analyser to be told to find a symbolic counterexample path
in which a cycle constraint $cc$ is satisfied by a symbolic state in the cycle part of the path,
and a global constraint $gc$ is satisfied by all symbolic states in the path.

In the fixpoint approach to symbolic LTL model checking,
the set of augmented states (states in the augmented model) that are fair with respect to $\varphi$ is computed.
A set of fairness constraints $C_{\varphi}$ is defined in a manner that is
dependent on the LTL encoding scheme. The fixpoint characterisation
of the set of augmented states that are fair with respect to $\varphi$ is
\begin{displaymath}
F_{\varphi} \triangleq \nu Z. \bigwedge_{c \in C_{\varphi}} r_B' (\mu Y. (Z \wedge c) \vee r_B'(Y))
\end{displaymath}
where $\nu$ and $\mu$ are respectively the greatest fixpoint and least fixpoint operators
of $\mu$-calculus \cite{koz83}.
The inner least fixpoint expression characterises states that satisfy $Z \wedge c$ directly
or do not satisfy $Z \wedge c$ but can reach states that satisfy $Z \wedge c$ in the augmented model,
where $Z$ is the variable of the outer greatest fixpoint operation.
The outer greatest fixpoint operation interacts with the inner fixpoint operations,
combining to ensure that each fairness constraint $c \in C_{\varphi}$ is satisfied infinitely often in each
path whose states are in $F_{\varphi}$.
(The overall fixpoint expression characterises states that can transition to states that can
start paths in which each $c \in C_{\varphi}$ is satisfied infinitely often.)

A symbolic augmented state $S_{\varphi}$, defined according to the LTL encoding scheme,
characterises the set of augmented states that
are ``commited'' to starting paths that satisfy $\varphi$. Let $I_{\varphi}$ denote the
symbolic augmented state characterising the set of initial augmented states.
The intersection characterised by
\begin{displaymath}
  I_{\varphi} \wedge S_{\varphi} \wedge F_{\varphi}
\end{displaymath}
determines if there is a counterexample for the LTL specification: the intersection is empty if and only if the
LTL specification has no counterexamples.

BT Analyser allows the computation of augmented states that are fair with respect
to $\varphi$ while satisfying a global constraint $gc$ globally. Note that in general
this is not equivalent to $F_{\varphi} \wedge gc$. Instead, the following fixpoint characterisation can be used:
\begin{equation}
\label{eq:gc}
\nu Z. \quad gc \wedge \bigwedge_{c \in C_{\varphi}} r_B' (\mu Y. (Z \wedge c) \vee r_B'(Y)).
\end{equation}
The intersection of the set of augmented states characterised by \eqref{eq:gc} with the set
characterised by $ I_{\varphi} \wedge S_{\varphi}$ determines if there is a counterexample path
for the LTL specification for which $gc$ holds globally.

In addition to the fixpoint approach to symbolic model checking, the model checker also
allows on-the-fly symbolic LTL model checking. The algorithm used is an adaptation of the
standard nested DFS algorithm \cite{cou92} with the following important differences:
\begin{itemize}
\item The adapted algorithm works with symbolic transitions (using elementary blocks) and symbolic states.
\item The adapted algorithm constructs the B\"uchi automaton implicitly and on-the-fly.
\end{itemize}
Details of the algorithm can be found in \cite{kro14}.

Finally, an important feature of BT Analyser is the ability to perform analysis incrementally.
As an example, an analysis can start with reachability analysis, followed by computation of fair states,
followed by computation of counterexample states (states that can be part of counterexample paths),
followed by the generation of a counterexample path. We can then tell BT Analyser
to find more counterexample paths as well as
fair states with global constraints without redoing the initial parts of the analysis.
This enables different strategies to be used for different problems.
An example is provided in \cite{kro14} where reachability analysis before model checking
is a good strategy for a prioritised BT model (where internal system transitions are prioritised over external
events) but is a bad strategy for a non-prioritised BT model.

\section{Generating Minimal Cut Sets}
\label{sec:generating}

We assume that component failure modes have been injected into the model.
An LTL specification whose violation represents a system failure is first model checked.
If a traditional top event represented by a state formula $top$ is to be used, then the LTL specification model checked would be
%\begin{displaymath}
 $\mathbf{G} \; \lnot top$.
%\end{displaymath}
Using our method, $\mathbf{G} \; \lnot top$ can be replaced by any LTL formula representing non-occurrence
of the behaviour associated with system failure.

In what follows let $\varphi$ denote the hazardous system behaviour being analysed (i.e., the negation of the LTL formula representing the desired system safety property).

Cut sets will be extracted from counterexample paths and verified to be minimal.
The entire process of generating all MCSs can be fully automated.

\subsection{Extracting Cut Sets from Counterexample Paths}
\label{sec:extracting}

Traditionally, a cut set is the set of basic events that occur in a behaviour leading to the top event.
Unfortunately, LTL does not include events. Even the more expressive temporal logic CTL* (see e.g., \cite{cla99}) does not include events.
The occurrence of an event is obviously irreversible: once an event occurs, from that point on, the
event has occurred. A simple way to encode the occurrence of an event is to use a Boolean
state variable, say $\mathit{eOccurred}$. Initially the value of $\mathit{eOccurred}$ is $\mathit{false}$.
When the event occurs, $\mathit{eOccurred}$ is set to $\mathit{true}$ and no behaviour can change it back to $\mathit{false}$.
This way of encoding the occurrence of an event is used by Lindsay {\em et al} in their method \cite{lin12b}.
We call the state variable for a basic event a {\em basic event flag}. 
A cut set can be encoded as a conjunction of the basic event flags
that correspond to elements of the cut set.

We can extract the cut set by choosing a symbolic state in the cycle part
and pick all positive occurrences of basic event flags.
As an example,
suppose there are  11 basic components that can fail
and their failure occurrences are represented by the basic event flags
\begin{displaymath}
\begin{array}{l}
\mathit{distyF, distgF, distbF, E1F, E2F, PTUF,}
\mathit{EDPyF, EDPgF, EMPbF, EMPyF,} \textrm{ and } \mathit{RATF.}
\end{array}
\end{displaymath}
We will denote the number of basic events $\mathit{MAX}$.
Thus for the above example $\mathit{MAX}=11$.
Suppose further that we choose the following symbolic state from the cycle part of the symbolic
counterexample path (taken from an actual run of the model checker on the A320 hydraulics case study):
\begin{displaymath}
\begin{array}{cl}
& \mathit{(Pilot = ready) \wedge (Engine1 = on) \wedge (Engine2 = off) \wedge (Yellow = off)} \\
\wedge & \mathit{(PTUy = off) \wedge (PTU = off) \wedge (EDPy = off) \wedge (EDPg = on) \wedge (Blue = off)} \\
\wedge & \mathit{(EMPy = off) \wedge (EMPb = off) \wedge  (RAT = off) \wedge \lnot E1F \wedge E2F \wedge \lnot distyF} \\
\wedge & \mathit{\lnot distgF \wedge \lnot distbF \wedge PTUF \wedge \lnot EDPyF
     \wedge \lnot EDPgF \wedge EMPbF \wedge EMPyF} \\
\wedge & \mathit{\lnot RATF \wedge (Green = on) \wedge (Aircraft = flyingSlow) \wedge (System = operating)},
\end{array}
\end{displaymath}
then the cut set extracted from the counterexample path would be
\begin{displaymath}
\mathit{\{E2F, PTUF, EMPbF, EMPyF}\}.
\end{displaymath}
Any symbolic state in the cycle can be chosen, thus we can always choose the starting symbolic state of the cycle
(basic event flags can only transition from $\mathit{false}$ to $\mathit{true}$, thus their values must remain
constant throughout a cycle).

If the symbolic state has a disjunction (i.e., the symbol ``$\vee$'' occurs in the proposition that is the symbolic state),
then the symbolic state can be normalised into an irredundant sum of product (ISOP) form
and a cut set can be extracted from any of the disjuncts. In BT Analyser, ISOP forms are
generated directly from OBDDs using the Minato-Morreale algorithm \cite{min93, mor70}.

\subsection{Verifying Minimality of Cut Sets}
\label{sec:verifying}

In isolation, if a cut set is extracted from a counterexample path, then
it can be checked for minimality by ensuring that there are no counterexample paths
if any member of the cut set is ``removed''.
This can be performed in the model checker by computing the set of states that are
fair with respect to $\varphi$ with a global constraint $gc$ representing the removal
of a member of the cut set. The computation is performed using \eqref{eq:gc}.
If the intersection of the resulting set of states with the
set of initial states is empty, then the cut set is minimal.
The idea goes as follows.
Continuing with the example from Section \ref{sec:extracting}, suppose the cut set
of interest (denoted $cs$) is
\begin{displaymath}
cs = \mathit{\{E2F, PTUF, EMPbF, EMPyF}\}.
\end{displaymath}
The $gc$ is a conjunction
of all the other components not failing together with at least one of the members
of the cut set also not failing, and for the above $cs$ we would have
\begin{displaymath}
\begin{array}{rcl}
gc & \equiv & \;\;\; \mathit{\lnot distyF \wedge \lnot distgF \wedge \lnot distbF \wedge  \lnot E1F \wedge \lnot EDPyF \wedge \lnot EDPgF \wedge \lnot RATF } \\
 & & \mathit{\wedge \; (\lnot E2F \vee \lnot PTUF \vee \lnot EMPbF \vee \lnot EMPyF)}.
\end{array}
\end{displaymath}
The conjunction $\mathit{\lnot distyF \wedge \lnot distgF \wedge \lnot distbF \wedge  \lnot E1F \wedge \lnot EDPyF
\wedge \lnot EDPgF \wedge \lnot RATF }$
represents components that have not failed and $\mathit{(\lnot E2F \vee \lnot PTUF \vee \lnot EMPbF \vee \lnot EMPyF)}$
represents the removal of at least one component from $cs$.
If the intersection of the result of \eqref{eq:gc} with the set of initial states is not empty,
then there is a counterexample path with global constraint $gc$,
meaning there is a proper subset of $cs$ that is a cut set,
and thus $cs$ is non-minimal. Otherwise $cs$ is minimal.

Computing the set of states that are fair with respect to $\varphi$ with global constraint $gc$ is usually
much faster than computing the set of states that are fair with respect to $\varphi$ without global constraints.
However, often this is still more expensive than counterexample path generation.
With some strategies to be discussed in Section \ref{sec:strategies}, minimality need not be checked
and the number of computations of fair states can be reduced.

\subsection{Strategies for Generating Minimal Cut Sets}
\label{sec:strategies}

Two different strategies for automating CSA are given here: a naive strategy and a systematic
strategy. The naive strategy finds MCSs in no specific order, while the
systematic strategy finds MCSs in a non-decreasing order in terms of size.

\subsubsection{Naive Strategy}

The steps of this strategy are as follows:
\begin{enumerate}
\item Compute the set of fair states.
\item Set $gc \gets true$ (i.e., no global constraints).
\item Find a counterexample path with global constraint $gc$ and no cycle constraints.
\item If a counterexample path is found, extract a cut set and verify its minimality as described in Section \ref{sec:verifying}.
If it is not minimal then find a counterexample path for the minimality and extract a smaller cut set.
This can be repeated until a minimal cut set is extracted. Add the MCS to the set of
MCSs sets found and modify $gc$ to rule out the found MCS from further consideration:
add the negation of a representation of the MCS as a conjunct to $gc$.
For the above example cut set, the negation is
\begin{displaymath}
\lnot(\mathit{E2F \wedge PTUF \wedge EMPbF \wedge EMPyF}).
\end{displaymath}
Repeat from step 3.
\item Otherwise no counterexample path was found that satisfies $gc$ globally
and all minimal cut sets have been found.
\end{enumerate}
A proof sketch of the correctness of the strategy is as follows:
\begin{itemize}
\item Each conjunct in $gc$ is associated with an MCS already found and rules out any  more cut sets that are subsumed
by the MCS.
\item In step 4, each time an MCS is found, an appropriate conjunct is added to $gc$.
\item Nowhere else is a conjunct added to $gc$.
\item If there are no counterexamples that satisfy $gc$
then there are no cut sets except those that are subsumed by the MCSs found in step 4
(since $gc$ rules out cut sets that are subsumed by the MCSs already found).
\end{itemize}
Since after step 5 we can conclude that all cut sets are subsumed by the MCSs found,
the MCSs found are exactly all the MCSs.

\subsubsection{Systematic Strategy}

As mentioned in Section \ref{sec:verifying}, verifying the minimality of a cut set by computing fair states with
global constraints can be expensive. A more systematic strategy that could be more
efficient would be to find all MCSs of
size 0, then all MCSs of size 1,
then all MCSs of size 2, and so on until we reach size $\mathit{MAX}$.
This strategy uses the cycle constraint feature of directed counterexample generation:
\begin{enumerate}
\item Compute the set of fair states.
\item Set $gc \gets true$ and $n \gets 0$.
\item Find a counterexample path with global constraint $gc$ and cycle constraint ``there are $n$ or less
basic events''. For example, suppose there are 3 basic events represented by $\mathit{C1F, C2F}$ and $\mathit{C3F}$.
For $n=1$ the cycle constraint would be
\begin{displaymath}
\mathit{ \lnot C1F \wedge \lnot C2F}
  \vee \mathit{ \lnot C1F \wedge \lnot C3F}
  \vee \mathit{ \lnot C2F \wedge \lnot C3F}.
\end{displaymath}
It may be easier to view the constraint as ``at least $\mathit{MAX} - n$ basic events have not happened''.
It is equivalent to the standard ``choosing $\mathit{MAX} - n$ out of $MAX$'' combinatorial problem.
\item If a counterexample path is found, the extracted cut set would be a minimal cut set.
Add the MCS to the set of
MCSs found and modify $gc$ to rule out the found MCS from further consideration
(as in step 4 of the naive strategy).
Repeat from step 3.
\item Otherwise no counterexample path was found that satisfies the constraint. If $n < \mathit{MAX}$,
set $n \gets n + 1$ and repeat from step 3. Otherwise $n \ge \mathit{MAX}$,
and all MCSs have been found.
\end{enumerate}
A proof sketch of the correctness of the strategy is as follows (it is slightly different than the one for the naive strategy):
\begin{itemize}
\item Each conjunct in $gc$ is associated with an MCS already found and rules out any  more cut sets that are subsumed
by the MCS.
\item Each time an MCS is found in step 3, an appropriate conjunct to rule out the MCS
found from further consideration is added to $gc$ in step 4.
\item Nowhere else is a conjunct added to $gc$.
\item If no more counterexample is found in step 3, then there are no cut sets of size $\le n$
except those subsumed by the MCSs already found.
\item A cut set found in step 3 is a minimal cut set:
the case where $n=0$ is trivial;
for $1 \le n \le \mathit{MAX}$, we can assume that all cut sets of size $\le n-1$
are subsumed by the MCSs already found
therefore any cut set found in step 3 is a minimal cut set
(since they are not subsumed by the MCSs already found,
thus they are not subsumed by any cut set of size $< n$).
\item If there are no counterexamples that satisfy $gc$ when $n = \mathit{MAX}$
then there are no cut sets of size $\le \mathit{MAX}$ except those that are subsumed by the MCSs already found.
\end{itemize}
Since the size of a cut set cannot be greater than $\mathit{MAX}$,
after step 5 with $n = \mathit{MAX}$ we can conclude that all cut sets are subsumed by the MCSs found,
thus the MCSs found are exactly all the MCSs.

\section{Experimental Results}
\label{sec:experiments}

Experiments were conducted to compare the performances of the two strategies described in
Section \ref{sec:strategies}. The naive strategy was then slightly modified, with on-the-fly symbolic
LTL model checking replacing directed counterexample path generation, and the performance
of the modified strategy compared with that of the original strategy.

\subsection{Comparison of Two Strategies}

The two strategies described in Section \ref{sec:strategies} were applied to the A320 hydraulics
case study used in \cite{bie02} and \cite{lin12b}. The BT model with the injected faults is
included as an example for BT Analyser, which can be accessed
using the DOI 10.14264/uql.2015.16
(it is part of the supplementary material for \cite{kro14}).
The experiments were performed on a notebook computer with a 2.7GHz i7 CPU and 16GB of RAM
running Ubuntu.
The BT model is similar to the one used in \cite{lin12b}.
The LTL specification used is
\begin{displaymath}
\mathbf{G}\Bigg(\lnot \mathit{INIT} \vee
\mathbf{G}\Bigg(\mathit{
\begin{array}{rl}
& (Yellow = on) \wedge (Green = on) \\
  \vee & (Yellow = on) \wedge (Blue = on) \\
\vee & (Green = on) \wedge (Blue = on)
\end{array}
}\Bigg)\Bigg)
\end{displaymath}
where $\mathbf{G}$ is the {\em globally} temporal operator ($\Box$ or box), and
\begin{displaymath}
 \mathit{INIT} \triangleq \mathit{(Yellow = on) \wedge (Green = on) \wedge (Blue = on)}.
\end{displaymath}
Both strategies produced the same set of MCSs,
although the order in which the MCSs were found were different.
Five 2-element MCSs were found:
\begin{displaymath}
\begin{array}{l}
\mathit{\{distyF, distbF\}, \{distyF, distgF\},}
\mathit{\{distgF, distbF\},} 
\mathit{\{distyF, EMPbF\}} \textrm{ and } \mathit{\{distgF, EMPbF\}.}
\end{array}
\end{displaymath}

There were ten 3-element MCSs found:
\begin{displaymath}
\begin{array}{l}
\mathit{\{distyF, PTUF, EDPgF\}, \{distbF, PTUF, EDPgF\}, \{distyF, E1F, PTUF\},} \\
\mathit{ \{distbF, E1F, PTUF\},}
\mathit{\{PTUF, EDPgF, EMPbF\}, \{E1F, PTUF, EMPbF\},} \\
\mathit{\{EDPyF, EDPgF, EMPyF\}, \{E1F, E2F, EMPyF\}, \{E2F, EDPgF, EMPyF\}} \\
\textrm{and } \mathit{\{E1F, EDPyF, EMPyF\}.}
\end{array}
\end{displaymath}

There were six 4-element MCSs found:
\begin{displaymath}
\begin{array}{l}
\mathit{\{distbF, PTUF, EDPyF, EMPyF\},}
\mathit{\{distgF, PTUF, EDPyF, EMPyF\},} \\
\mathit{\{distbF, E2F, PTUF, EMPyF\},}
\mathit{\{distgF, E2F, PTUF, EMPyF\},} \\
\mathit{\{PTUF, EDPyF, EMPbF, EMPyF\}} \textrm{ and }
\mathit{\{E2F, PTUF, EMPbF, EMPyF\}.}
\end{array}
\end{displaymath}

Table \ref{tbl:timing} shows the overall timing results for the two strategies.
Both strategies have the same initial model checking time for the LTL specification
which is 8 minutes. The subsequent MCS generation times differed substantially.

\begin{table}
\centering
\caption{A320 Hydraulics Timing Results}
\label{tbl:timing}
\begin{tabular}{|c|c|c|c|}
\hline
 Strategy & Initial MC run & MCS Generation & Total   \\
\hline\hline
Naive & 8 minutes & 57 minutes & 65 minutes \\
\hline
Systematic & 8 minutes & 5 minutes & 13 minutes \\
\hline
\end{tabular}
\end{table}

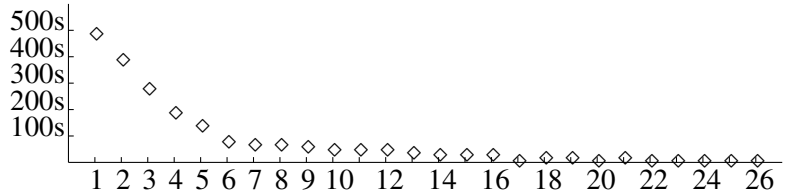
\begin{figure}[h]

\centering
\begin{picture}(300,60)

\put(-22,10){100s}
\put(0,10){\line(1,0){2}}
\put(-22,20){200s}
\put(0,20){\line(1,0){2}}
\put(-22,30){300s}
\put(0,30){\line(1,0){2}}
\put(-22,40){400s}
\put(0,40){\line(1,0){2}}
\put(-22,50){500s}
\put(0,50){\line(1,0){2}}

\put(0,0){\line(1,0){270}}
\put(0,0){\line(0,1){60}}

\put(8,46){$\diamond$}
\put(10,0){\line(0,1){2}}
\put(8,-10){1}
\put(18,36){$\diamond$}
\put(20,0){\line(0,1){2}}
\put(18,-10){2}
\put(28,25){$\diamond$}
\put(30,0){\line(0,1){2}}
\put(28,-10){3}
\put(38,16){$\diamond$}
\put(40,0){\line(0,1){2}}
\put(38,-10){4}
\put(48,11){$\diamond$}
\put(50,0){\line(0,1){2}}
\put(48,-10){5}
\put(58,5){$\diamond$}
\put(60,0){\line(0,1){2}}
\put(58,-10){6}
\put(68,4){$\diamond$}
\put(70,0){\line(0,1){2}}
\put(68,-10){7}
\put(78,4){$\diamond$}
\put(80,0){\line(0,1){2}}
\put(78,-10){8}
\put(88,3){$\diamond$}
\put(90,0){\line(0,1){2}}
\put(88,-10){9}
\put(98,2){$\diamond$}
\put(100,0){\line(0,1){2}}
\put(97,-10){10}
\put(108,2){$\diamond$}
\put(110,0){\line(0,1){2}}
%\put(110,-10){11}
\put(118,2){$\diamond$}
\put(120,0){\line(0,1){2}}
\put(116,-10){12}
\put(128,1){$\diamond$}
\put(130,0){\line(0,1){2}}
%\put(130,-10){13}
\put(138,0){$\diamond$}
\put(140,0){\line(0,1){2}}
\put(136,-10){14}
\put(148,0){$\diamond$}
\put(150,0){\line(0,1){2}}
%\put(150,-10){15}
\put(158,0){$\diamond$}
\put(160,0){\line(0,1){2}}
\put(156,-10){16}
\put(168,-2){$\diamond$}
\put(170,0){\line(0,1){2}}
%\put(170,-10){17}
\put(178,-1){$\diamond$}
\put(180,0){\line(0,1){2}}
\put(176,-10){18}
\put(188,-1){$\diamond$}
\put(190,0){\line(0,1){2}}
%\put(190,-10){19}
\put(198,-2){$\diamond$}
\put(200,0){\line(0,1){2}}
\put(196,-10){20}
\put(208,-1){$\diamond$}
\put(210,0){\line(0,1){2}}
%\put(210,-10){21}
\put(218,-2){$\diamond$}
\put(220,0){\line(0,1){2}}
\put(216,-10){22}
\put(228,-2){$\diamond$}
\put(230,0){\line(0,1){2}}
%put(230,-10){23}
\put(238,-2){$\diamond$}
\put(240,0){\line(0,1){2}}
\put(236,-10){24}
\put(248,-2){$\diamond$}
\put(250,0){\line(0,1){2}}
%\put(250,-10){25}
\put(258,-2){$\diamond$}
\put(260,0){\line(0,1){2}}
\put(256,-10){26}

\end{picture}
\caption{CTR Generation Time for Naive Strategy (26 counterexample paths)}
\label{fig:str1a}
\end{figure}

\begin{figure}[h]

\centering
\begin{picture}(320,60)

\put(-12,10){2s}
\put(0,10){\line(1,0){2}}
\put(-12,20){4s}
\put(0,20){\line(1,0){2}}
\put(-12,30){6s}
\put(0,30){\line(1,0){2}}
\put(-12,40){8s}
\put(0,40){\line(1,0){2}}
\put(-16,50){10s}
\put(0,50){\line(1,0){2}}

\put(0,0){\line(1,0){320}}
\put(0,0){\line(0,1){60}}

\put(8,20){$\diamond$}
\put(10,0){\line(0,1){2}}
\put(8,-10){1}
\put(18,13){$\diamond$}
\put(20,0){\line(0,1){2}}
\put(18,-10){2}
\put(28,21){$\diamond$}
\put(30,0){\line(0,1){2}}
\put(28,-10){3}
\put(38,21){$\diamond$}
\put(40,0){\line(0,1){2}}
\put(38,-10){4}
\put(48,22){$\diamond$}
\put(50,0){\line(0,1){2}}
\put(48,-10){5}
\put(58,20){$\diamond$}
\put(60,0){\line(0,1){2}}
\put(58,-10){6}
\put(68,19){$\diamond$}
\put(70,0){\line(0,1){2}}
\put(68,-10){7}
\put(78,20){$\diamond$}
\put(80,0){\line(0,1){2}}
\put(78,-10){8}
\put(88,32){$\diamond$}
\put(90,0){\line(0,1){2}}
\put(88,-10){9}
\put(98,35){$\diamond$}
\put(100,0){\line(0,1){2}}
\put(97,-10){10}
\put(108,32){$\diamond$}
\put(110,0){\line(0,1){2}}
%\put(110,-10){11}
\put(118,32){$\diamond$}
\put(120,0){\line(0,1){2}}
\put(116,-10){12}
\put(128,26){$\diamond$}
\put(130,0){\line(0,1){2}}
%\put(130,-10){13}
\put(138,41){$\diamond$}
\put(140,0){\line(0,1){2}}
\put(136,-10){14}
\put(148,27){$\diamond$}
\put(150,0){\line(0,1){2}}
%\put(150,-10){15}
\put(158,21){$\diamond$}
\put(160,0){\line(0,1){2}}
\put(156,-10){16}
\put(168,23){$\diamond$}
\put(170,0){\line(0,1){2}}
%\put(170,-10){17}
\put(178,29){$\diamond$}
\put(180,0){\line(0,1){2}}
\put(176,-10){18}
\put(188,22){$\diamond$}
\put(190,0){\line(0,1){2}}
%\put(190,-10){19}
\put(198,36){$\diamond$}
\put(200,0){\line(0,1){2}}
\put(196,-10){20}
\put(208,33){$\diamond$}
\put(210,0){\line(0,1){2}}
%\put(210,-10){21}
\put(218,40){$\diamond$}
\put(220,0){\line(0,1){2}}
\put(216,-10){22}
\put(228,31){$\diamond$}
\put(230,0){\line(0,1){2}}
%put(230,-10){23}
\put(238,30){$\diamond$}
\put(240,0){\line(0,1){2}}
\put(236,-10){24}
\put(248,24){$\diamond$}
\put(250,0){\line(0,1){2}}
%\put(250,-10){25}
\put(258,31){$\diamond$}
\put(260,0){\line(0,1){2}}
\put(256,-10){26}
\put(268,34){$\diamond$}
\put(270,0){\line(0,1){2}}
%\put(266,-10){27}
\put(278,33){$\diamond$}
\put(280,0){\line(0,1){2}}
\put(276,-10){28}
\put(288,23){$\diamond$}
\put(290,0){\line(0,1){2}}
%\put(286,-10){29}
\put(298,33){$\diamond$}
\put(300,0){\line(0,1){2}}
\put(296,-10){30}
\put(308,34){$\diamond$}
\put(310,0){\line(0,1){2}}
%\put(306,-10){31}
\put(318,24){$\diamond$}
\put(320,0){\line(0,1){2}}
\put(316,-10){32}

\end{picture}
\caption{CTR Generation Time for Systematic Strategy (32 counterexample paths)}
\label{fig:str2}
\end{figure}
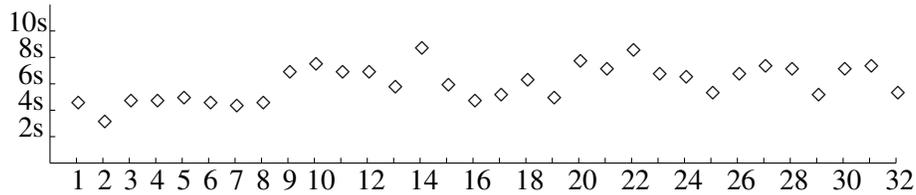

The use of cycle constraints drastically reduced the counterexample generation time.
As Fig. \ref{fig:str1a} shows, the counterexample generation times for the naive strategy
started at a high of 490 seconds and dropped significantly as more global constraints were added.
In contrast, the counterexample generation times for the systematic strategy (which uses
cycle constraints) did not vary as much and were quick at
between 3 and 9 seconds (see Fig. \ref{fig:str2}).

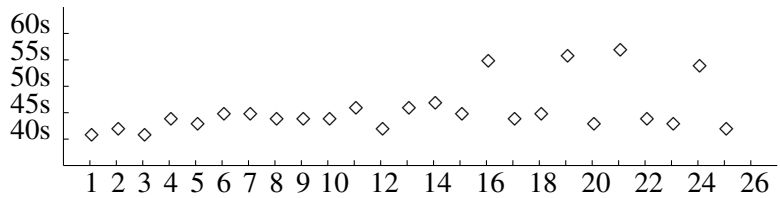
\begin{figure}[h]

\centering
\begin{picture}(300,60)

\put(-20,10){40s}
\put(0,10){\line(1,0){2}}
\put(-20,20){45s}
\put(0,20){\line(1,0){2}}
\put(-20,30){50s}
\put(0,30){\line(1,0){2}}
\put(-20,40){55s}
\put(0,40){\line(1,0){2}}
\put(-20,50){60s}
\put(0,50){\line(1,0){2}}

\put(0,0){\line(1,0){270}}
\put(0,0){\line(0,1){60}}

\put(8,9){$\diamond$}
\put(10,0){\line(0,1){2}}
\put(8,-10){1}
\put(18,11){$\diamond$}
\put(20,0){\line(0,1){2}}
\put(18,-10){2}
\put(28,9){$\diamond$}
\put(30,0){\line(0,1){2}}
\put(28,-10){3}
\put(38,15){$\diamond$}
\put(40,0){\line(0,1){2}}
\put(38,-10){4}
\put(48,13){$\diamond$}
\put(50,0){\line(0,1){2}}
\put(48,-10){5}
\put(58,17){$\diamond$}
\put(60,0){\line(0,1){2}}
\put(58,-10){6}
\put(68,17){$\diamond$}
\put(70,0){\line(0,1){2}}
\put(68,-10){7}
\put(78,15){$\diamond$}
\put(80,0){\line(0,1){2}}
\put(78,-10){8}
\put(88,15){$\diamond$}
\put(90,0){\line(0,1){2}}
\put(88,-10){9}
\put(98,15){$\diamond$}
\put(100,0){\line(0,1){2}}
\put(97,-10){10}
\put(108,19){$\diamond$}
\put(110,0){\line(0,1){2}}
%\put(110,-10){11}
\put(118,11){$\diamond$}
\put(120,0){\line(0,1){2}}
\put(116,-10){12}
\put(128,19){$\diamond$}
\put(130,0){\line(0,1){2}}
%\put(130,-10){13}
\put(138,21){$\diamond$}
\put(140,0){\line(0,1){2}}
\put(136,-10){14}
\put(148,17){$\diamond$}
\put(150,0){\line(0,1){2}}
%\put(150,-10){15}
\put(158,37){$\diamond$}
\put(160,0){\line(0,1){2}}
\put(156,-10){16}
\put(168,15){$\diamond$}
\put(170,0){\line(0,1){2}}
%\put(170,-10){17}
\put(178,17){$\diamond$}
\put(180,0){\line(0,1){2}}
\put(176,-10){18}
\put(188,39){$\diamond$}
\put(190,0){\line(0,1){2}}
%\put(190,-10){19}
\put(198,13){$\diamond$}
\put(200,0){\line(0,1){2}}
\put(196,-10){20}
\put(208,41){$\diamond$}
\put(210,0){\line(0,1){2}}
%\put(210,-10){21}
\put(218,15){$\diamond$}
\put(220,0){\line(0,1){2}}
\put(216,-10){22}
\put(228,13){$\diamond$}
\put(230,0){\line(0,1){2}}
%put(230,-10){23}
\put(238,35){$\diamond$}
\put(240,0){\line(0,1){2}}
\put(236,-10){24}
\put(248,11){$\diamond$}
\put(250,0){\line(0,1){2}}
%\put(250,-10){25}
%\put(258,-2){$\diamond$}
\put(260,0){\line(0,1){2}}
\put(256,-10){26}

\end{picture}
\caption{Minimality Checking Time for Naive Strategy (25 iterations)}
\label{fig:str1b}
\end{figure}

The times for computing fair states with global constraints to check minimality did not vary so much,
as can be seen in Fig. \ref{fig:str1b}. They were between 40 and 60 seconds.
In fact, the four cases where the times were significantly higher were exactly
for cases where the cut set turned out to be non-minimal: cases 16, 19, 21 and 24.
Interestingly, the performance penalty for those cases were somewhat offset by
the significantly faster counterexample generation times that immediately followed:
cases 17, 20, 22 and 25 in Fig. \ref{fig:str1a}.
There is no case 26 for checking minimality since no counterexample was found
for the case.

The results from the experiments clearly show that cycle constraints
in directed counterexample generation can drastically reduce the computation time
required to find counterexample paths.
Moreover, cycle constraints can be used to direct the counterexample path
generation towards specific kinds of paths. For generating MCSs using
the systematic strategy, the cycle constraint directs the model checker
to find a counterexample path that produces a cut set of exactly the desired size.

The results also show the utility of global constraints. Global constraints
serve two purposes:
\begin{itemize}
\item
to rule out certain classes of counterexamples, and
\item
to reduce the search space for counterexamples.
\end{itemize}
The second is a pleasant consequence of the first.
Generally, a smaller search space results in faster search time.
The effect is rather dramatic in counterexample generation for the naive strategy (Fig. \ref{fig:str1a}).

Finally, the results show the benefits of incremental analysis described in Section \ref{sec:mc}.
Even a bad strategy (the naive strategy) is helped by incremental analysis.
The computation time using the bad strategy is still better than the total computation
time, using SAL (rerun on the same notebook computer used for the experiments), of the original approach in \cite{lin12b} (1 hour vs 4 hours).
Although there are too many factors that are not taken into account for a fair comparison,
it is clear that not redoing work can save a lot of time.

\subsection{Using On-the-fly Symbolic LTL Model Checking}

Recall from Section \ref{sec:mc} that on-the-fly symbolic LTL model checking can be used to find counterexample paths
quickly. However, the on-the-fly symbolic LTL model checking facility in BT Analyser is not as well developed as
directed counterexample path generation. In particular the facility does not yet support cycle constraints.
As a result, the facility cannot be used for the systematic strategy.

An experiment was conducted with on-the-fly symbolic LTL model checking replacing directed counterexample
path generation for the naive strategy. In the experiment, the existence of any remaining counterexample path
(and thus any remaining MCSs)
was checked first in each round of MCS generation. This is because on-the-fly symbolic LTL checking might take
a long time if there are no counterexamples.

\begin{table}
\centering
\caption{Results for Naive Strategy with Symbolic On-the-fly LTL Checking}
\label{tbl:otf}
\begin{tabular}{|c|c|c|c|c|}
\hline
 $|ICS|$ & $|MCS|$ & CTR Existence Test & Total OTF & Total Minimality Test   \\
\hline\hline
4 & 3 & 415.5s & 6.8s & 141s \\
\hline
6 & 3 & 414.0s & 21.4s & 246.7s  \\
\hline
9 & 2 & 321.1s & 72.2s & 833.0s \\
\hline
8 & 2 & 271.8s & 39.8s & 738.1s \\
\hline
9 & 3 & 239.8s & 81.8s & 812.8s \\
\hline
8 & 2 & 346.7s & 38.6s & 781.6s \\
\hline
5 & 3 & 262.5s & 23.3s & 194.1s \\
\hline
5 & 2 & 254.3s & 25.5s & 361.7s \\
\hline
5 & 3 & 244.3s & 23.5s & 266.0s \\
\hline
7 & 2 & 230.6s & 33.4s & 615.0s \\
\hline
5 & 3 & 206.3s & 24.6s & 272.6s \\
\hline
5 & 3 & 210.2s & 27.5s & 196.9s \\
\hline
5 & 3 & 201.1s & 27.9s & 268.8s \\
\hline
6 & 3 & 191.6s & 28.9s & 385.9s \\
\hline
5 & 4 & 192.8s & 22.2s & 166.4s \\
\hline
6 & 4 & 190.0s & 30.4s & 279.2s \\
\hline
6 & 4 & 193.1s & 28.2s & 282.1s \\
\hline
4 & 3 & 191.2s & 28.9s & 128.6s \\
\hline
4 & 4 & 169.3s & 26.4s & 57.8s \\
\hline
5 & 4 & 164.7s & 31.3s & 167.5s \\
\hline
5 & 4 & 164.6s & 33.4s & 164.3s \\
\hline
 & & 98.1s & & \\
\hline
\hline
 & & 5173.6s & 676.0s & 7360.1s \\
\hline
\end{tabular}
\end{table}

Table \ref{tbl:otf} shows the results of the experiment with on-the-fly symbolic LTL checking.
For each round (represented by a row), the column $|ICS|$ is for the size of the initial cut set found,
the column $|MCS|$ is for the size of the MCS found, the column ``CTR Existence Test'' is
for the time it took to check the existence of a counterexample path (by computing fair states),
the column ``Total OTF'' is for the total of the on-the-fly symbolic LTL checking times, and
the column ``Total Minimality Test'' is for the total of the minimality test times.
The number of iterations for a round is $|ICS| - |MCS| + 1$, thus
for $|ICS| = 9$ and $|MCS| = 2$, there would have been 8 iterations of
on-the-fly symbolic LTL checking and minimality test in the round.
The second last row represents the time needed to verify that there are no more MCSs.

The results show that, although the total of on-the-fly symbolic LTL checking times (11 minutes) is much less
than the total of the counterexample generation times for the naive strategy with directed
counterexample path generation (57 minutes), the overall
analysis time is significantly greater. Even if we ignore the existence test times for the rounds
and use 98.1 seconds as  the time to check the absence of
more counterexamples (we can imagine a parallelised implementation where on-the-fly symbolic LTL
checking is run in parallel with checking the absence of counterexamples and whichever result is
obtained first is used, with the other computation aborted), the total time is still around 135 minutes,
compared to 65 minutes using directed counterexample path generation.

The disappointing performance when using on-the-fly symbolic LTL checking for the naive strategy appears
to be caused by the {\em depth-first-search with imprecise (but safe) tracking of fairness constraints}
nature of the algorithm used. This resulted in counterexample paths that tended to produce cut sets
with many more failed components than necessary. Of the 21 MCSs, only one of them
was found in one iteration, with the other 20 MCSs requiring 78 iterations in all.
(By contrast, with directed counterexample path generation, 17 of the MCSs
were found in one iteration and the remaining 4 were found in two iterations.)
Thus, any savings from the fast counterexample generation times were completely wiped out
by the times required for the minimality tests.

Perhaps if cycle constraints can be incorporated into on-the-fly symbolic LTL checking,
then the systematic strategy might benefit from fast counterexample path generation times
of on-the-fly symbolic LTL checking without the initial overhead of computing fair states.

\section{Discussion}
\label{sec:dis}

The approach described above, and BT Analyser, were developed in a very general typed multi-variable transition system modelling framework. The results were illustrated on a translation from the Behavior Tree notation into the framework but could easily be extended to support other modelling notations, including state-based notations such as VDM, Z and B. The main restriction is that all sets must be bounded finite sets and the number of threads is bounded.
Note that if the model has a terminating state, then the state must be replaced by one that has a ``null'' transition (i.e., it can
transition to itself). This ensures that all paths are infinite.

Fault injection in BT models is usually very straightforward. Although they are all in some sense equivalent, the BT model of a system can have different tree shapes depending on the order in which the subtrees corresponding to different requirements were integrated (see \cite{dro03} for a description of the BT model development process). If they have been developed so that individual component behaviours are grouped together by component, which is very often the case, then fault injection is simply a matter of grafting a behaviour tree corresponding to the component failure mode into the tree, with the failure (basic) event at the graft point. 

Failure Modes and Effects Analysis (FMEA) is a commonly used consequence analysis technique for critical systems. It consists of checking whether individual component failures can have hazardous effects on the system. Tool support can be provided for FMEA by injecting individual faults into system models and using model checking to see if undesirable system states are reachable. Grunske {\em et al} \cite{gru11} reports experience automating FMEA using BT models and the SAL model checker. Although in principle this approach could be extended to CSA simply by injecting multiple faults into the model, one for each element of a potential cut set, in practice the limits of model checking are quickly reached due to the state explosion problem. By contrast, BT Analyser can handle system models like those in \cite{gru11} in which all of the faults are injected, and it does so very efficiently.

The generalisation of {\em top event} from a failure state to a behaviour allows the approach to be applied
to a broad class of modelling notations, including notations that do not explicitly represent state changes, such as the BT notation.

\section{Related Work}
\label{sec:relwork}

Bieber {\em et al} \cite{bie02} combine CSA and model checking in safety analysis of the A320 hydraulics case study.
They use two different models, one of the static, physical architecture and an LTL model of the activation and
failure behaviour of the system, and need reasoning to combine the results to prove the safety requirements, whereas our approach uses a single model and the entire analysis is automated.

Akerlund {\em et al} \cite{ake99} propose that reliability analysis such as FTA and verification of safety requirements
be performed on the same model. Their generation of MCSs uses the reachability analysis
capability of a model checker, but is based on a state condition as the top event.
The FSAP/NuSMV-SA safety analysis platform \cite{boz07} and the DCCA method \cite{ort06} also use the reachability analysis
capabilities of model checkers to generate MCSs based on a state condition.
In contrast, our approach is applicable when the top event is a behaviour.

Abdulla {\em et al} \cite{abdulla06} use a SAT-based model checker to find MCSs for Scade models, in which component failure modes are modelled by replacing nominal flows by extended flows.
Again, as with the approaches mentioned in the previous paragraph, they perform reachability analysis to
a state condition.

Papadopoulos and Maruhn \cite{papadopoulosMaruhn01} construct a fault tree from the design model. Ortmeier and Schellhorn \cite{OrtSch07} model the fault trees themselves and verify their
correctness and completeness against Statechart models. Cha {\em et al} \cite{ChaSon03} use the model checker UPPAAL to verify the correctness of fault trees against timed-automata models.
In contrast, our approach does not need to involve fault trees.
 
Lindsay {\em et al} \cite{lin12b} use a model checker to generate MCSs from failure behaviour.
However, some of the steps still need to be performed manually.

\section{Summary and Conclusions}
\label{sec:concl}

We have presented an approach that can be used to generate minimal cut sets automatically from a fault-injected
model and a general safety requirement.
The approach takes advantage of novel features of a new model checker, BT Analyser. In particular, the approach
uses the incremental analysis capability of BT Analyser, directed counterexample
path generation, cycle constraints and global constraints.
The use of these features significantly reduces the computation time required when compared to
the traditional way of using model checkers.

Experiments on an existing case study in the Behavior Tree notation show the effectiveness of the approach and the
novel features of BT Analyser. The effectiveness of cycle constraints in
producing counterexample paths with desired properties and in producing them quickly
was a pleasant surprise. The use of constraints appears to help mitigate the state explosion
problem in model checking.

\section{Future Work}
\label{sec:future}

The on-the-fly symbolic LTL model checking facility in BT Analyser is not as developed as
directed counterexample path generation. In particular, cycle constraints cannot be specified
for counterexample paths. A possible future work would be to incorporate cycle constraints
into the on-the-fly symbolic LTL checking mechanism.

Another possible future work is to incorporate partial order reduction techniques \cite{pel96} into
on-the-fly symbolic LTL model checking in BT Analyser. This may speed-up on-the-fly checking in cases
where there are no counterexamples.

\section{Acknowledgements}

This work was supported by Linkage Project grants LP0989363 and LP130100201 from the Australian Research
Council, Raytheon Australia and Thales Australia.

\nocite{*}
\bibliographystyle{eptcs}
\bibliography{references}

\end{document}